# A Study on the Relevance of Information in Discriminative and Non-Discriminative Media

Carlos Gershenson[1], Mason A. Porter[2], Andrej Probst[3], Matus Marko[4], Atin Das[5]

[1]School of Cognitive and Computer Sciences, University of Sussex, Brighton, BN1 9QN, U. K.
C.Gershenson@sussex.ac.uk. [2]Center for Applied Mathematics, Cornell University.
[3]Faculty of Mathematics, Physics, and Informatics, Comenius University, Bratislava, Slovakia.
[4]Faculty of Management, Comenius University, Bratislava, Slovakia.
[5]Department of Mathematics, Jadavpur University, Calcutta, India.

*Abstract*

*In this paper we compare the relevance of information obtained from "discriminative" media and from "non-discriminative" media. Discriminative media are the ones which accumulate and deliver information using a heuristic selection of it. This can be made by humans, or by artificial intelligent systems, exhibiting some form of "knowledge". Non-discriminative media just collect and return information without any distinction. This can also be made by humans or by artificial systems, but there is no "knowledge" involved in the process.*

*We ranked the words occurring in an edited electronic publication specialized in complex systems research, and we found that they approximate a modified Zipf distribution. We compared occurrences of representative words from the distribution with the occurrences in non-discriminative media. We found that a non-discriminative medium (Google) has a higher variance from of our original distribution than a semi-discriminative one (NEC Research Index), even when both appear to have their own modified Zipf distribution.*

*We conclude that discriminative media have a higher efficiency rating, at least in the area in which they specialize, than non-discriminative media. Using the same search method, the discriminative media should deliver more relevant information. This relevancy also depends on the skills of the user, but non-discriminative media are more sensitive to poor searching skills, as there is a higher probability of delivering irrelevant information. This leads us to suggest the incorporation of intelligent classifications in different media (such as the ones suggested by the Semantic Web project), in order to increase the relevance of the delivered information.*

## 1. Introduction

We can discriminate information systems depending on the *relevance* of the information provided by them. If the information is relevant for our purposes, we can say that the system is *useful* for us, as we will need to spend less time and effort in extracting the information we need. We could also say that this information has a higher quality if it is relevant for our purposes than irrelevant information. Notice that the relevance is always tied to a context, so we cannot speak about relevance of information of a system without referring to the use of the information we will give. Therefore, if we would like to define an utility function of the information provided by a system, we should take into account its relevance, the context for which this information is relevant, the effort required by the user to obtain relevant information, and the probability of retrieving relevant information.

We will label *discriminative media* the ones which obtain and return relevant information for a specific context. For this to be done, there needs to be a heuristic, as the system *appears* to "know" which information is relevant and which is not. On the other hand, *non-discriminative media* will be the ones which receive and deliver "a bulk" of information, leaving the user with the task of filtering the relevant information. In a discriminative medium, this filtering can be made by humans, who enter the information (already discriminated) to the medium, or processed by "intelligent" software (Das *et al.*, 2002).

In order to observe how the quality of the information varies according to the degree of discrimination of a medium, we study word occurrences from The Complexity Digest (www.comdig.org), an edited electronic publication in the field of complexity. These are presented in the following section. Then, we compare the obtained word occurrences with information from NEC Research Index (citeseer.nj.nec.com/cs), and Google (www.google.com). The results are shown in Section 3. We conclude suggesting the incorporation of "intelligent" software into different media for increasing the utility of the information they deliver.

## 2. A Discriminative Medium

In order to study the distribution of a specialized medium, we ranked the word occurrences from the Complexity Digest (1999-2001). The Complexity Digest consists of a weekly distribution of excerpts or abstracts from articles related to complexity, found on the Internet by the editors. We used a "stoplist", in order not to include prepositions, articles, or pronouns, such as "from", "a", "the","he", etc. The plot of the first four thousand ranked words can be seen in Figure 1.

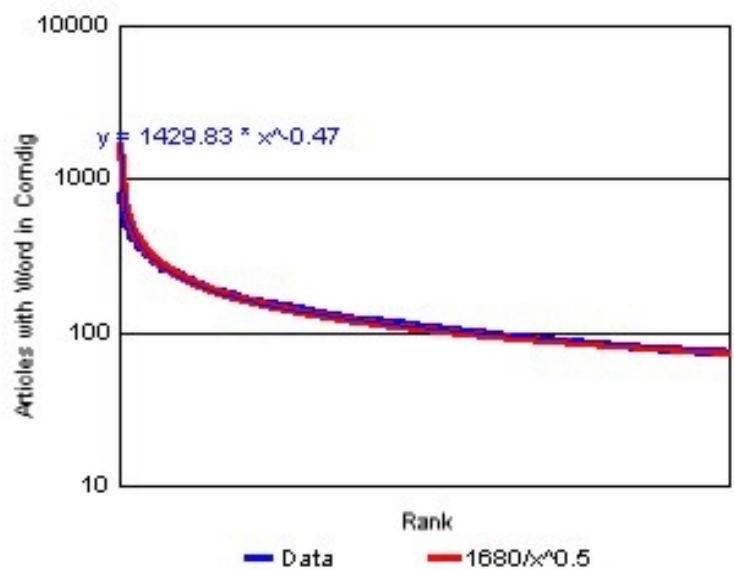

Figure 1. Word occurrence and modified Zipf approximation.

We plotted the rank of the words against the number of occurrences per article, and found that they match a common log fit ($y=1429.83x^{-0.47}$), but also a modified Zipf distribution ($y=1680/x^{0.5}$) (Zipf, 1949; Gell-Mann, 1994). The constants in the chart for the modified Zipf distribution were chosen by inspection of the data. An optimal fit would yield even better results. The fact that the rank of the words follow a Zipf distribution indicates that they also fit a Pareto distribution and a power law distribution (Adamic, 2000).

There is no complete understanding in the meaning of something having a Zipf distribution, since so many data sets have a Zipf or power-law distribution (Gell-Mann, 1994; Barabási and Albert, 1999), which are equivalent between them and also with the Pareto distribution (Adamic, 2000). What is clear is that there are few words repeated many times, and many words repeated few times. And among the small group of frequent words, we find many words pertinent for complexity (See Table 1). The obvious reason for this is that the editors chose articles relevant in complexity, where these words are abundant.

Since the Complexity Digest is focussed in the study of complexity, it will have a wider distribution of information related to this field. Another reason is that it is a specialized medium, but specialized media can be considered as a special case of discriminative media, as they discriminate information into the area they specialize in. Nevertheless, there can be general discriminative media. It matters that they deliver relevant information related to a context, but they can deliver relevant information related to one or few contexts (specialized), or to several (general). The three media studied here deliver information with similar search methods, so increasingly specialized media become more discriminative, and — vice versa — increasingly general media become less discriminative.

## 3. Comparison with Non-Discriminative Media

We performed searches in NEC Research Index and Google of some relevant words we found in Complexity Digest. Even when these media have a modified Zipf distribution fitting their ranking of word occurrence, the ranking of the words are expected to be very different from that of Complexity Digest, as the first one deals with research in all areas, not only complexity, and Google delivers information from all over the Internet. Therefore, the latter is more general than the former.

Figure 2 shows the ranking of the most common seventy words in Complexity Digest, and a common log fitting their distribution. Figure 3 shows representative words in the same order for NEC Research Index, and Figure 4 shows the same information for Google. Table 1 shows some examples of words ranked in Complexity Digest and their occurrences in the three media. It is apparent that words related to complexity are ranked much higher in the Complexity Digest, which is a discriminative medium, as it is edited by humans who decide which articles are relevant. On Google, as it is a much more general medium, and thus non-discriminative, the number of pages with information relevant to complexity, which are found by the representative words, is much lower in comparison to the information delivered from other contexts. This is also because the information of Google is obtained without any discrimination. NEC Research Index lies somewhere between these two extremes, as much of the terminology is common with that of complexity, but since there are also many other research areas not related to complexity, we can see that the information has some probability of being less relevant than in a discriminative medium, such as the Complexity Digest. There is a discriminative criterium here, as it accepts only research publications, but it is more general than the Complexity

Digest. Also, as Google is a more general medium, the variation of the word occurrences is much higher than in NEC Research Index, in comparison to the Complexity Digest.

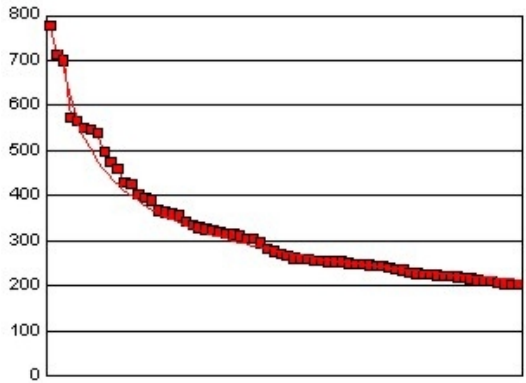

Figure 2. Words with highest ranks in the Complexity Digest.

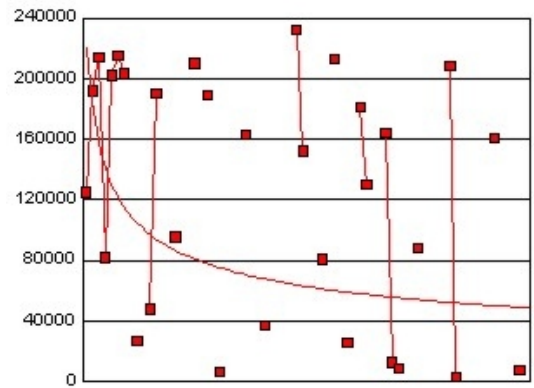

Figure 3. Word occurrence for NEC Research Index

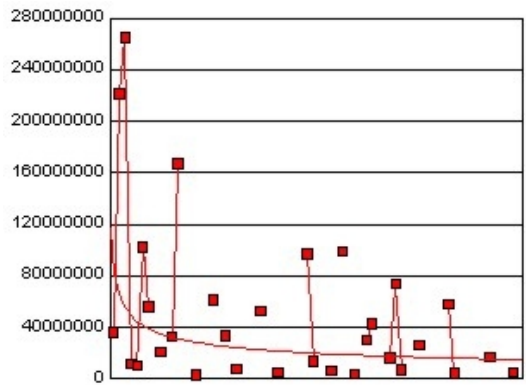

Figure 4. Word occurrence for Google. Same ordering as previous two.

| Rank | Word | Complexity Digest | NEC Research Index | Google |
|---|---|---|---|---|
| 1 | science | 777 (100.00) | 125,000 (100.00) | 36,200,000 (100.00) |
| 4 | complex | 573 (73.75) | 81,900 (65.52) | 11,400,000 (31.49) |
| 6 | system | 550 (70.79) | 215,000 (172.00) | 102,000,000 (281.77) |
| 11 | human | 461 (59.33) | 47,800 (38.24) | 32,800,000 (90.61) |
| 12 | time | 429 (55.21) | 190,000 (152.00) | 167,000,000 (461.33) |
| 18 | research | 363 (46.72) | 210,000 (168.00) | 61,300,000 (169.34) |
| 22 | brain | 336 (43.24) | 6,580 (5.26) | 7,930,000 (21.91) |
| 38 | evolution | 259 (33.33) | 80,700 (64.56) | 6,310,000 (17.43) |
| 58 | computer | 224 (28.83) | 208,000 (166.40) | 58,200,000 (160.77) |
| 1087 | chemistry | 40 (5.15) | 2,310 (1.85) | 6,330,000 (17.49) |
| 3837 | door | 10 (1.29) | 845 (0.68) | 22,100,000 (61.05) |

Table 1. Comparison of word occurrence in different media (normalized at 100 for occurrences of "science").

Recall that we are addressing the relevance of information. Because the information delivered by Google is much larger than that by NEC Research Index, and this one is much larger than the one of the Complexity Digest, one expects to find by amount more information about complex systems first in Google, then in NEC Research Index, and then in the Complexity Digest. Very probably this is so, but we are discussing the amount of relevant information delivered by the medium in relation to the amount of irrelevant information ("signal-to-noise" ratio). In other words, using the same search methods and knowledge of the user, it is much easier to obtain irrelevant information to "complexity" in Google than in NEC Research Index than in the Complexity Digest. An experienced user can likely obtain relevant information from a general medium without many problems, but then the "knowledge" for discriminating successfully is not in the medium, but in the user. A user who knows nothing about "complexity" aside from its name is served best by searching in the most discriminative medium available.

We can see that word occurrences are one way for determining the relevance of information. That is to say, if we identify relevant words in a specific area, information which contains a high frequency of these words will very probably be relevant for that area. This can be similar to the development of an *ontology*. In computer science, an ontology can be seen as a formal and explicit specification of a shared conceptualization (Marko *et al.*, 2002). An ontology is useful for determining the use and meaning of words in specific contexts. We believe that the analysis of word frequencies can be helpful for developing ontologies, both manually and automatically. And vice versa: an ontology can determine which words should be searched for a high frequency in order to detect relevant information.

## 4. Conclusions

We can see that an important part in determining the degree of discrimination of a medium is the success in which it delivers relevant contextual information. One way to do this is by structuring the database where the information is stored, so that only the relevant one is retrieved at the necessary moment. A complementary way is to do "intelligent" searches for the relevant information. An effort for achieving both of these is being carried out by the Semantic Web project (World Wide Web Consortium, www.w3.org/2001/SW; also see Marko *et al.*, 2002). This includes the creation of *ontologies*, which determine the context to which some information is related, and eases the way in which this information can be processed and combined with other sources. Ontologies can be defined by humans, but there is also an effort in automatizing the process of ontology creation.

These developments in information technology should improve the relevance of the information obtained using automated media, making them more discriminative, and therefore more useful.

## 5. Acknowledgements

The authors would like to thank the Complexity Digest (www.comdig.org) for providing the data analysed in this work and Gottfried Mayer-Kreiss for valuable suggestions and comments. C. G. was partly supported by the Consejo Nacional de Ciencia y Tecnología (CONACYT) of México and by the School of Cognitive and Computer Sciences of the University of Sussex.